\documentstyle[aps,prl,manuscript]{revtex}
\begin{document}

\draft

%\twocolumn[\hsize\textwidth\columnwidth\hsize\csname
%@twocolumnfalse\endcsname

\title{Zero-bias anomaly in one-dimensional tunneling contacts}
\bigskip

\author{Michael Reizer}

\address{5614 Naiche Rd. Columbus, OH 43213} 

\maketitle

%\widetext
\begin{abstract}
We study the Coulomb interaction effects on the tunneling 
conductance of a contact constructed of two parallel
quantum wires. The following contacts are considered:
two clean identical quantum wires, two disordered identical
quantum wires, and asymmetric contact of one clean and
another disordered quantum wires. We show that the low-voltage anomaly of 
the tunneling conductance is less singular than the low-energy
anomaly of the one-particle density of states.

 \end{abstract}
 \bigskip

 \pacs{PACS numbers: 71.45.Gm, 73.40.Gk}
 
% ]
% \narrowtext
 %2col
\bigskip
\centerline{\bf 1. Introduction}
\bigskip
Transport of interacting electrons in quantum wires was considered in 
Ref. \cite{GRS} for the Wigner crystal, and in Refs. \cite{KF} 
and \cite{MG} for
one  and multi-channel Lattinger liquid. The approach incorporating the
Landauer formalism with the description of the electron-electron interaction
as interaction of an electron with fluctuating electromagnetic field
created by the other electrons was developed in Refs \cite{O}.
The results of these works were applied to quantum wires
with constriction or with an impurity. Low and high potential
barriers were considered.

In the present paper we consider a different situation, the tunneling
transport between two parallel quantum wires. Zero-bias anomaly in the
one-particle density of states of disordered electron systems including
the quasi-one-dimensional case is well known \cite{AA}.
Zero-bias anomaly of the tunneling conductance of disordered
two-dimensional contacts were studied in Refs. \cite{N} and \cite{LS} 
applying the tunneling action method (quasiclassical approach).
Alternative approach developed in Refs. \cite{KR} and \cite{R},
based on the tunneling hamiltonian and the linear response method, 
naturally includes the interlayer and intra-layer
electron-electron interaction. This method 
makes a direct connection between collective excitation in 
a two-layer electron system and the tunneling anomaly and it allows
to consider on equal footing disordered systems where collective excitations
are diffusion modes and clean systems where collective excitations are
gapless two-dimensional plasmons. The asymmetric 
clean-disordered contact was considered in Ref. \cite{R}.    
In the present paper we extend the analysis of Refs. \cite{KR} and \cite{R}
for one dimension.

Plan of the paper is the following. We start with the first order
Coulomb electron-electron interaction correction to one-particle
density of states and compare it with the result obtained earlier by
the bosonization technique \cite{TA}. Then we study the tunneling
conductance of symmetric clean or disordered quantum wires and 
compare the anomaly in the tunneling conductance with corresponding
anomalies of the one-particle density of states. Finally we consider 
an asymmetric contact of one clean and another disordered quantum wires.
\bigskip
 
\centerline{\bf 2. Density of states of one-dimensional} 
\centerline{\bf interacting electron systems}
\bigskip

The one-particle electron density of states is defined as
\begin{equation}
\label{1}
\nu(\epsilon)=-{2\over \pi}\int{dp\over 2\pi}{\rm Im}[G^R(P)],     
\end{equation}
where $G^R(P)$ is the retarded electron Green's functions.
Without the electron-electron interaction
the electron Green's function is  
\begin{equation}
\label{2}
G^R_0(P)=[G^A_0(P)]^*={1\over \epsilon-\xi_p+i/2\tau},\ \ \ \xi_p={p^2-p^2_F
\over 2m},                            
\end{equation}
where $P=({\bf p},\epsilon)$, and $\tau$ is the
elastic electron-impurity relaxation time, in this chapter we assume
$1/\tau=0$
The correction to electron density of states in one dimension due to 
the Coulomb interaction is
\begin{equation}
\label{3}
\delta\nu(\epsilon)=-{2\over \pi}{\rm Im}\int{dp\over 2\pi}[G^R(P)]^2
\Sigma_{e-e}^R(P),
\end{equation}
where the electron self-energy is
\begin{eqnarray}
\label{4}
\Sigma_{e-e}^R(P)=\int{dQ\over (2\pi)^2}\biggl[{\rm Im}[G_0^A(P+Q)]
V^A(Q)\tanh\biggl({\epsilon+\omega\over 2T}\biggr)\cr 
+G^R(P+Q){\rm Im}V^A(Q)\coth\biggl({\omega)\over 2T}\biggr)\biggr],
\end{eqnarray}
where $Q=({\bf q},\omega)$.  

The screened one-dimensional Coulomb potential is 
\begin{equation}
\label{5}
V^A(Q)={V_0(q)\over 1-V_0(q)P^A(Q)}.                
\end{equation}
The nonscreened Coulomb potential $V_0(q)$ is
\begin{equation}
\label{6}
V_0(q)={2e^2\over\epsilon_0}\ln\biggl({1\over qa}\biggr),
\ \ \ qa<<1,
\end{equation}                                                        
where $a$ is a width of a wire
and $\epsilon_0$ is the static dielectric constant, which we will absorb into
$e^2$.
The polarization operator in the long-wave-length limit, $\omega>>qv_F$, is 
\begin{equation}
\label{7}
P^A(Q)={2v_Fq^2\over \pi(\omega-i0)^2}.                         
\end{equation}
which lead to one-dimensional plasmons $\omega_q^2=(2/\pi)V_0(q)v_Fq^2$.

Integrating over the electron momentum in Eq. (3) we have 
\begin{eqnarray}
\label{8}
{\delta\nu(\epsilon, T)\over \nu}=-{1\over (2\pi)^2}\int_{-\infty}^\infty
d\omega \tanh\biggl({\epsilon+\omega\over 2T}\biggr)
{\rm Im}\int_{-\infty}^\infty dq
{V_0(q)\over (\omega-i0)^2-(2/\pi)V_0(q)v_Fq^2}.                             
\end{eqnarray}
where $\nu=1/(\pi v_F)$.
Performing momentum integration  within the logarithmic accuracy we get,
\begin{eqnarray}
\label{9}
{\delta\nu(\epsilon, T)\over \nu}
={1\over 4}\biggl({e^2\over \pi v_F}\biggr)^{1/2}
\int_{-\infty}^\infty {d\omega\over \omega} 
\tanh\biggl({\epsilon+\omega\over 2T}\biggr)
\times[\ln(E/|\omega|)]^{1/2},\ \ \
E=\biggl({2e^2v_F\over\pi a^2}\biggr)^{1/2}
\end{eqnarray}
The correction to the density of states for $T<<\epsilon$ is
\begin{equation}
\label{10}
{\delta\nu(\epsilon, T)\over \nu}=
-{1\over 6}\biggl({e^2\over \pi v_F}\biggr)^{1/2}
[\ln(E/\epsilon)]^{3/2},        
\end{equation}
Equation (10) corresponds to the first term in the expansion of
the equation
\begin{equation}
\label{11}
\nu(\epsilon)={1\over \pi v_F}\exp\biggl[
-{1\over 6}\biggl({e^2\over \pi v_F}\biggr)^{1/2}
[\ln(E/\epsilon)]^{3/2}\biggr].                    
\end{equation}
obtained earlier by bosonization technique 
in Ref. \cite{TA}.  

\bigskip

\centerline{\bf 2. Two clean quantum wires}
\bigskip
The tunneling current in the linear response theory is given by
\begin{equation}
\label{12}j=-e{\rm Im}\Pi^R(eV),                                    
\end{equation}
where $e$ is the electron charge, $V$ is the applied voltage, and 
$\Pi^R$ is the retarded polarization operator with lower and upper Green's 
functions related to different banks, and the tunneling matrix element $t$ 
stands as vertices. The momentum
does not conserve at vertices. For noninteracting electrons 
the tunneling conductance of asymmetric contact is  
\begin{equation}
\label{13}
G_0={\partial j\over \partial V}={1\over 2}e^2\pi|t|^2\nu_1\nu_2, 
\end{equation}
where $\nu_1$ and $\nu_2$ are the electron density of states in each
bank of the tunneling contact. We assume
that both banks have the same chemical potentials to ensure that
we have zero-bias anomaly, thus $\nu_1=\nu_2=\nu$.
 
Now we discuss the Coulomb interaction in a system of two
parallel identical quantum wires.
The nonscreened Coulomb potentials within the wire, $V_0$, and 
between electrons in different wires, $U_0$, are 
\begin{eqnarray}
\label{14}
V_0(q)=2e^2\ln(1/qa),\ \ \ qa<<1,\cr 
U_0(q)=2e^2\ln(1/qd),\ \ \ qd<<1, 
\end{eqnarray}                          
where $d$ is the distance between the wires.
The screened interlayer $U$ and
intra-layer potentials $V$ are described by the equations
\begin{eqnarray}
\label{15}
U=U_0+V_0P U+U_0P V, \cr
 V=V_0+V_0P V+U_0P U,          
\end{eqnarray}
where $P$ is the polarization operator defined by Eq. (16).
It is convenient to present the potentials $V$ and $U$ in the form
\begin{eqnarray}
\label{16}
V={1\over 2}{V_0+U_0\over 1-P(V_0+U_0)}
+{1\over 2}{V_0-U_0\over 1-P(V_0-U_0)},    
\end{eqnarray}
\begin{eqnarray}
\label{17}              
U={1\over 2}{V_0+U_0\over 1-P(V_0+U_0)}
-{1\over 2}{V_0-U_0\over 1-P(V_0-U_0)}.                    
\end{eqnarray}
In all further calculations small momentum transfers are important, thus we 
assume $qa<<1$ and  $qd<<1$ thus
\begin{eqnarray}
\label{18}
V_0-U_0=(1/2)\pi v_FA,\ \
V_0^2-U_0^2=2e^2A\pi v_F\ln(1/qb),
b=(ad)^{1/2},\ \ A={4e^2\over \pi v_F}\ln(d/a).
\end{eqnarray}
The contribution from the intra-layer and inter-layer interactions are
\begin{eqnarray}
\label{19}
\delta_{11}G+\delta_{22}G
=e^2|t|^2{\rm Im}\int{d\epsilon\over 2\pi}
\int{d\omega\over 2\pi}{\partial S(\epsilon+eV)\over \partial\epsilon}
\times S(\epsilon+\omega)\cr
\int{dq\over 2\pi}V^R(Q)
\int{dp'\over 2\pi}({\cal G}^A-{\cal G}^R)(p',\epsilon)
\times\int{dp\over 2\pi}({\cal G}^A(P))^2{\cal G}^R(P+Q),
\end{eqnarray}
\begin{eqnarray}
\label{20}
\delta_{12}G=e^2|t|^2{\rm Im}\int{d\epsilon\over 2\pi}
\int{d\omega\over 2\pi}
{\partial S(\epsilon+eV)\over \partial\epsilon}S(\epsilon+\omega)\cr
\times\int{dq\over 2\pi}U^R(Q)
\biggl[\int{dp\over 2\pi}{\cal G}^A(P){\cal G}^R(P+Q)\biggr]^2,
\end{eqnarray}
where $S(\epsilon)=-\tanh(\epsilon/2T)$.
 
The combined contribution after integrating over electron momentums 
in the region $qv_F<|\omega|$ is
\begin{eqnarray}
\label{21}
{\delta G\over G_0}={1\over G_0}\sum_{ij}\delta_{ij}G
=-{1\over \pi^2}\int d\omega f(\omega,eV)J_c(\omega),\cr
\end{eqnarray}
where
\begin{eqnarray}
\label{22}
J_c(\omega)={\rm Im}\int_0^{|\omega|/v_F} 
dq{V^R(q,\omega)-U^R(q,\omega)\over \omega^2},                   
\end{eqnarray}
\begin{eqnarray}
\label{23}
f(\omega,eV)={1\over 2}\int{d\epsilon}
{\partial S(\epsilon+eV)\over \partial\epsilon}S(\epsilon+\omega).
\end{eqnarray}
We see that the result depends on the effective potential $V-U$.
According to Eqs. (16-18),
\begin{eqnarray}
\label{24}
V^R-U^R={V_0-U_0\over 1-P^R(V_0-U_0)}=
{1\over 2}{A\pi v_F\omega^2\over (\omega+i0)^2-Av_Fq^2}.  
\end{eqnarray}
If $A>>1$ the pole in Eq. (24) is within the limit of
integration of Eq. (22) and 
\begin{eqnarray}
\label{25}
J_c(\omega)=-{\pi^2\over 4\omega}A^{1/2}.
\end{eqnarray}
Further we consider the case of $T<<eV$, thus
\begin{eqnarray}
\label{26}
{\delta G\over G_0}={A^{1/2}\over 4}\ln(eV/E).  
\end{eqnarray}
We see that singularity in the tunneling conductance is weaker than
in the one-particle density of states because the effective potential
$V-U$ is less singular than $V$.

\bigskip
\centerline{\bf 2. Two disordered quantum wires}
\bigskip
We consider the tunneling contact of two identical disordered 
quantum wires. The disordered quantum wire is considered 
as a quasi-one-dimensional system for voltages satisfying 
the condition $eV<<D/a^2$, see Ref. \cite{AA}, where $D$ is the 
diffusion coefficient.
The polarization operator in a disordered wire is
\begin{eqnarray}
\label{27}
P^R(q,\omega)=-\nu{Dq^2\over -i\omega+Dq^2},\ \ \    q\ell<<1,
\ \ \ \omega\tau<<1,                                             
\end{eqnarray}
where $\ell=v_F\tau$ is the electron mean free path. 
Generalizing corresponding equation of Ref. 
\cite{R} for the quasi-one-dimensional case
we get
\begin{eqnarray}
\label{28}
{\delta G\over G_0}=-{1\over 2\pi(p_Fa)^2}\int d\omega f(\omega,eV)J_d(\omega),
\end{eqnarray}
where
\begin{eqnarray}
\label{29}
J_d(\omega)={\rm Im}\int_0^{1/\ell} 
dq{V^R(Q)-U^R(Q)\over (-i\omega+Dq^2)^2}.  
\end{eqnarray}
The effective potential $V-U$ is
\begin{eqnarray}
\label{30}
V^R(Q)-U^R(Q)=2e^2\ln(d/a){-i\omega+Dq^2\over -i\omega+D'q^2},\ \   
D'=(1+A/2)D.                     
\end{eqnarray}
Therefore
\begin{eqnarray}
\label{31}
J_d(\omega)=-{\pi^2 v_FA\over (2\omega)^{3/2}}
{1\over (D')^{1/2}-D^{1/2}}.                     
\end{eqnarray}
Finally we have
\begin{eqnarray}
\label{32}
{\delta G\over G_0}=-{1\over 4}
[(1+A/2)^{1/2}+1]
\biggl({1\over 2eV\tau}\biggr)^{1/2},
\end{eqnarray}
for 
\begin{eqnarray}
\label{33}
{1\over \tau(p_Fa)^4}<<eV<<{\rm min}\biggl[{1\over \tau}, {D\over a^2}\biggr].
\end{eqnarray}
The left inequality garantees applicatibility of the perturbation theory
for Eq. (32).

Comparing Eq. (32) with the correction to the one-particle
density of states in disordered quasi-one-dimensional case \cite{AA}
\begin{eqnarray}
\label{34}
\delta\nu(\epsilon)\sim{1\over \epsilon^{1/2}(p_Fa)^2}
(\ln\epsilon)^{1/2},                                              
\end{eqnarray}
we again see that the correction to the tunneling conductance is
less singular than the density of states.

\bigskip
\centerline{\bf 3. One clean and another disordered quantum wires}
\bigskip

The screened potentials are satisfied the equations
\begin{eqnarray}
\label{35}
\pmatrix{V_{11}&U_{12}\cr U_{21}&V_{22}}=
  \pmatrix{V_0&U_0\cr U_0&V_0}                      
+\pmatrix{V_0&U_0\cr U_0&V_0}\pmatrix{P_1&0\cr 0&P_2}
\pmatrix{V_{11}&U_{12}\cr U_{21}&V_{22}}.
\end{eqnarray}
We will use the definitions: $V_{11}=V_1$, $V_{22}=V_2$, and
$U_{12}=U_{21}=U$.
The solution of Eq. (35) is
\begin{eqnarray}
\label{36}
U={U_0\over {\cal D}},\ \ 
V_{1,2}={V_0-(V_0^2-U_0^2)P_{2,1}\over {\cal D}},\cr 
{\cal D}=(1-V_0P_1)(1-V_0P_2)-U_0^2P_1P_2.                        
\end{eqnarray}
We assume layer 1 is clean and layer 2 is disordered, 
The polarization operator in the disordered wire, $P_2$, is defined by
Eq. (27) and $P_1$ by Eq. (7) for $qv_F<<\omega$ and $P_1=-\nu$ for
$\omega<<qv_F$.
The corrections to the tunneling conductance in an asymmetric two-dimensional
contact were derived in Ref. \cite{R}. 
Generalizing them for the one-dimensional
case we have
\begin{eqnarray}
\label{37}
{\delta_{22} G\over G_0}=-{1\over 2\pi(p_Fa)^2}
\int d\omega f(\omega,eV)
\times{\rm Im}\int_0^{1/\ell'} dq{V_2^R(Q)\over (-i\omega+Dq^2)^2},
\end{eqnarray}
\begin{eqnarray}
\label{38}
{\delta_{11}G\over G_0}={1\over 2\pi(p_Fa)^2}\int d\omega f(\omega,eV)
{\rm Im}\int_0^{|\omega|/v_F}dq{V_1^R(Q)\over \omega^2}, 
\end{eqnarray}
\begin{eqnarray}
\label{39}
{\delta_{12}G\over G_0}={1\over \pi^2(p_Fa)^2}\int d\omega  f(\omega,eV)
\times{\rm Im}\int_0^{1/\ell'} dq{\Gamma^R(Q)U^R(Q)
\over -i\omega+Dq^2}.
\end{eqnarray}                                           
where
\begin{eqnarray}
\label{40}
\Gamma^R(Q)={1\over \nu}\int {dp\over 2\pi}
{\cal G}^A(P){\cal G}^R(P+Q)
={i\pi\over \omega-qv_F},
\end{eqnarray}                                                                 
and $\ell'={\rm max}[\ell, a, d]$.  
There are two characteristic regions of momentum in Eqs. (37-39),
the plasmon region, $q<|\omega|/v_F<1/\ell'$, and the diffusion region,
$|\omega|/v_F<q<1/\ell'$.
In the plasmon region $Dq^2<<|\omega|$ and
 $\Gamma^R(q,\omega)\approx i\pi/\omega$, thus the combined
contribution from all terms in Eqs. (37-39) is 
\begin{eqnarray}
\label{41}
{\delta G\over G_0}=
{1\over 2\pi(p_Fa)^2}\int d\omega f(\omega,eV)J(\omega),
\end{eqnarray}
where
\begin{eqnarray}
\label{42}
J(\omega)={\rm Im}\int_0^{|\omega|/v_F}{dq\over \omega^2}
(V_1^R(Q)+V_2^R(Q)-2U^R(Q)).                 
\end{eqnarray}
In the plasmon region $(qv_F/\omega)^2>>Dq^2/|\omega|$, 
thus $|P_1|>>|P_2|$, and the effective potential $V_1+V_2-2U$ takes the form 
\begin{eqnarray}
\label{43}
V_1+V_2-2U={1\over {\cal D}}[2(V_0-U_0)-(V_0^2-U_0^2)(P_1+P_2)]\cr
={\pi v_FA\biggl[
1-\frac{\displaystyle 2e^2}{\displaystyle \pi v_F}
\frac{\displaystyle (qv_F)^2 }{\displaystyle \omega^2} 
\ln(1/qb)\biggr]\over 
1-\frac{\displaystyle 2e^2}{\displaystyle \pi v_F}
\frac{\displaystyle(qv_F)^2 }{\displaystyle \omega^2} 
\biggl[\ln(1/qa)+A
\frac{\displaystyle Dq^2}{\displaystyle -i\omega}\ln(1/qb)\biggr]},
\end{eqnarray}
It is clear from Eqs. (42) and (43) that the important
contribution to the tunneling conductance appears only if 
\begin{eqnarray}
\label{44}
\ln(1/qa)<<A{Dq^2\over |\omega|}\ln(1/qb).
\end{eqnarray}
Inequality (44) sets the lower limit $q_0$ in the integral of Eq. (42).
Within logarithmic accuracy $q_0=(|\omega|/2AD)^{1/2}$. 
Provided inequality (44) is satisfied, the singular contribution to
the tunneling conductance originates from the following part of the
effective potential 
\begin{eqnarray}
\label{45}
V_1+V_2-2U
=-{2iAe^2\omega(qv_F)^2\ln(1/qb)
\over i\omega^3+2A(e^2/\pi v_F)Dq^2
(qv_F)^2\ln(1/qb)}. 
\end{eqnarray}
If besides inequality $\omega\tau<<1$ we assume also that $A\omega\tau<<1$,
then we can neglect $i\omega^3$ term in the denominator of Eq. 45.
As a result 
\begin{eqnarray}
\label{46}
J(\omega)=-{\pi v_F\over \omega^{3/2}}\biggl({A\over D}\biggr)^{1/2}, 
\end{eqnarray}
and the correction to the conductance is
\begin{eqnarray}
\label{47}
{\delta G\over G_0}={1\over 2(p_Fa)^2}\biggl({3A\over eV\tau}\biggr)^{1/2},
\end{eqnarray}
for
\begin{eqnarray}
\label{48}
{1\over \tau(p_Fa)^4}<<eV<<{\rm min}\biggl[{1\over A\tau}, 
{D\over a^2}\biggr].
\end{eqnarray}
Now we consider the diffusion region where only to terms $\delta_{22}G$
and $\delta_{12}G$ contribute. In the diffusion region 
$\omega\sim Dq^2$, and $\omega<<qv_F$, thus
following Eq. (19) we can put $P_1=-\nu$, and present the function 
${\cal D}(Q)$ in the following form
\begin{eqnarray}
\label{49}
(-i\omega+Dq^2){\cal D}(Q)
=(1+V_0\nu)(-i\omega+D''q^2),\ \
{D''\over D}=1+{V_0\nu+(V_0^2-U_0^2)\nu^2\over 1+V_0\nu}
\end{eqnarray}

We start with the correction $\delta_{22}G$. The momentum integral in Eq. (37)
\begin{eqnarray}
\label{50}
J_1(\omega)={\rm Im}\int_{|\omega|/v_F}^{1/\ell'}dq{V_2^R(Q)
\over (-i\omega+Dq^2)^2}                        
\end{eqnarray}
will be calculated within logarithmic accuracy while the lower limit
will be shifted to zero and the upper one to infinity.
As a result
\begin{eqnarray}
\label{51}
J_1(\omega)={\pi\over \omega^{3/2}(2D)^{1/2}}
[1+C(\omega)],
\end{eqnarray}
where
\begin{eqnarray}
\label{52}
C^2(\omega)={2e^2\nu(\ln(1/q_1a)+A\ln(1/q_1b))\over 1+2e^2\nu\ln(1/q_1a)},
\end{eqnarray}
where $q_1=(\omega/D)^{1/2}$.
Finally for the voltage satisfying condition of Eq. (33) we have
\begin{eqnarray}
\label{53}
{\delta_{22}G\over G_0}=
-{1\over 2(p_Fa)^2}[1+C(eV)]\biggl({1\over 2eV\tau}\biggr)^{1/2}.
\end{eqnarray}

Calculating $\delta_{12}G$ we take into account 
that in the diffusion region 
 $\Gamma^R(q,\omega)=-i\pi/qv_F$, thus 
\begin{eqnarray}
\label{54}
{\delta_{12}G\over G_0}=
-{1\over \pi^2v_F}\int d\omega f(\omega,eV)J_2(\omega),\cr
\end{eqnarray}
where
\begin{eqnarray}
\label{55}
J_2(\omega)={\rm Re}\int_0^{1/\ell'} dq
{U^R(Q)\over -i\omega+Dq^2}.                            
\end{eqnarray}
Using Eq. (36) for the potential $U$ we get 
\begin{eqnarray}
\label{56}
J_2(\omega)=
{\rm Re}\int_0^{1/\ell'}{dq\over-i\omega +D''q^2}{2\pi e^2\ln(1/qd)
\over 1+2\pi e^2\nu\ln(1/q_1d)}\cr
={\pi\over 4\omega}{1\over D''(q_1)}{2e^2\ln(1/q_1d)
\over 1+2e^2\nu\ln(1/q_1d)}
\end{eqnarray}
We see that $J_2(\omega)$ is less singular than $J_1(\omega)$ 
and therefore  the correction $\delta_{12}G$ in the diffusion region
may be neglected.                       
\bigskip

\centerline{\bf 4. Conclusions}
\bigskip
We considered the following types of tunneling contacts: 
two clean identical quantum wires, two disordered identical quantum wires,
and an asymmetric contact of one clean and another disordered
quantum wires.
We have shown that in symmetric contacts the zero-bias anomaly 
of the tunneling conductance due to the Coulomb
interaction is weaker than corresponding singularity of the
one-particle density of states. This effect is due to partial
cancellation of contributions from intra-wire and inter-wire interactions.
In an asymmetric clean-disordered contact only the anomaly
associated with the disordered wire survived, though slightly modified.
We see the zero-bias anomaly as a signature of collective excitations,
such as plasmons in clean double-wire systems or diffusion modes in
disordered double-wire system. Absence of a plasmon mode in the
coupled clean-disordered system lead to absence of corresponding
anomaly in the tunneling conductance.

The author is grateful to I. L. Aleiner, B. L. Altshuler, and D. L. Maslov
for useful discussions.


\bigskip

\begin{references}
\bibitem{GRS} L. I. Glazman, I. M. Rudin, and B. I. Shklovskii, Phys. Rev. B,
{\bf 45}, 8454 (1992).
\bibitem{KF} C. L. Kane and M. P. A. Fisher, Phys. Rev. Lett. {\bf 68},
1220 (1992).
\bibitem{MG} K. A. Matveev and L. I. Glazman, Phys. Rev. Lett. {\bf 70},
990 (1993).
\bibitem{O} A. A. Odintsov, Physica B {\bf 189}, 258 (1993).
\bibitem{AA} B. L. Altshuler and A. G. Aronov, {\it Electron-electron
interaction in disordered systems}, edited by A. L. Efros and
M. Polak, North-Holland, Amsterdam, (1985).
\bibitem{N} Y. Nazarov, Sov. Phys. JETP {\bf 68}, 561 (1989);
Sov. Solid St. Phys.{\bf 31}, 1581 (1990).
\bibitem{LS} L.Levitov and A. Shitov, in {\it Correlated Fermions and
Transport in Mesoscopic Systems}, p. 513, Editions Frontieres (1996). 
\bibitem{KR} D. V. Khveshchenko and M. Reizer, Phys. Rev. B {\bf 57},
4245 (1998).
\bibitem{R} M. Reizer,  Phys. Rev. B {\bf 58}, 15789 (1998).
\bibitem{TA} A. M. Tsvelik and B. L. Altshuler, (1992) unpublished.
\end{references}
\end{document}